\documentclass[aps,pre,twocolumn,preprintnumbers,amsmath,amssymb,superscriptaddress]{revtex4}
\usepackage{graphicx,color} 
\usepackage{dcolumn}
\usepackage{bm}
\usepackage{lipsum} 
\usepackage{subfigure}
\usepackage{relsize}

\newcommand \beq {\begin{equation}}
\newcommand \eeq {\end{equation}}
\newcommand \ben {\begin{eqnarray}}
\newcommand \een {\end{eqnarray}}

\begin{document}

\title{Phase Field Crystal Model for Magneto-Elasticity in Isotropic Ferromagnetic Solids} 

\author{Niloufar Faghihi}
\affiliation{Department of Applied Mathematics, The University of Western
Ontario, 1151 Richmond St. N., London, Ontario, Canada N6A 5B7}
\affiliation{Department of Physics, McGill University, 3600 rue University,
Montr\'{e}al, Qu\'{e}bec, Canada H3A 2T8}

\author{Nikolas Provatas} \affiliation{Department of Physics, McGill
University, 3600 rue University, Montr\'{e}al, Qu\'{e}bec, Canada H3A 2T8}

\author{K. R. Elder} \affiliation{Department of Physics, Oakland University,
Rochester, Michigan 48309, USA}

\author{Martin Grant} \affiliation{Department of Physics, McGill University,
3600 rue University, Montr\'{e}al, Qu\'{e}bec, Canada H3A 2T8}

\author{Mikko Karttunen}
\affiliation{Department of Applied Mathematics, The University of Western
Ontario, 1151 Richmond St. N., London, Ontario, Canada N6A 5B7}
\affiliation{Department of Chemistry \& Waterloo Institute for Nanotechnology, 
University of Waterloo, 200 University Avenue West, Waterloo, Ontario, Canada N2L 3G1}

\begin{abstract}

A new isotropic magneto-elastic phase field crystal (PFC) model 
to study the relation between morphological structure 
and magnetic properties of pure ferromagnetic solids
is introduced.  Analytic calculations were used 
to determine the phase diagram and obtain 
the relationship between elastic strains and magnetization. Time 
dependent numerical simulations were used to demonstrate the effect of
grain boundaries on the formation of magnetic domains. It was shown that
the grain boundaries act as nucleating sites for domains of reverse magnetization.
Finally, we derive a relation for coercivity versus grain mis-orientation in the isotropic limit.  
\end{abstract}

\date{\today}

\maketitle

\section{Introduction}

The physical properties of materials and their functions are often influenced 
by their microstructures \cite{mchenry-99,mchenry-00,provatas-07}.
This holds true in magnetic materials where for example, 
the magnetic coercivity, remanence and magnetic saturation are known to be 
a strong function of grain size in polycrystalline materials.  For 
applications it often useful to be able to taylor specific magnetic properties.  
For example in applications such as in magnetic data storage 
devices, sensors, motors, generators, transformers \cite{white-85} it desirable 
to have soft magnetic materials which have high large saturation and remanent 
magnetization and low coercivity.  These properties make them suitable for 
electronic devices for which a quick change of magnetization with minimum energy loss 
per cycle is required.   Thus it is very important to understand the 
detailed and complex relationship between microstructures and magnetization. 

Anisotropy is one of the key properties of magnetic materials.
To reverse the direction of a magnetic domain, a magnetic 
anisotropy energy barrier must be overcome such that the magnetic moments have enough energy to deviate 
from the easy axis of magnetization. The other factor that determines coercivity is the 
local morphology of the material. For example, grain boundaries and other defects 
can modify the barrier for formation of a magnetic domain \cite{fukunaga-92,rave-97,skomski-01}. 
Another example is a local composition variation: a secondary soft magnetic phase could affect 
the initiation of magnetization reversal.
Finally, the exchange interaction and magnetic dipole interactions also influence coercivity. 
These interactions establish collective alignment of magnetic moments which facilitates 
the formation of the magnetic domains \cite{fukunaga-92,hernando-00}.

Understanding the relation of magnetic microstructure to crystalline
microstructure can be useful for designing materials with desired magnetic
properties.  This is particularly true of nanocrystalline materials with grain
sizes on the order of a few tens of nanometers, where grain size and magnetic
correlation lengths are comparable. The relative magnitude of these length
scales can alter the mechanism of magnetic domain formation to be controlled
either by a long-range cooperative magnetic behaviour or by a local ordering.
This gives rise to a relatively broad range of coercivity values attainable in
nanocrystalline magnetic materials \cite{herzer-97} and opens up new
opportunities to develop ultra-high-density magnetic storage devices
\cite{scherfl-00}. Understanding the physics of these systems, however,
requires a robust modeling formalism that is capable of describing
magneto-elastic interactions in the presence of crystalline defects and
microstructural processes that evolve on diffusive time scales. 

One of the first studies of the effect of crystallographic structure on the magnetic properties
was by Harris \textit{et al.} \cite{zuckermann-73}. They suggested a
simple Hamiltonian to describe magnetism in amorphous materials with random
anisotropy. Later, this model was developed to study amorphous materials in
which the magnetic correlation length spans many length scales with different
anisotropy directions \cite{alben-becker-chi-78, chi-alben-77}:
For small grains the magnetic correlation length is larger than the grain size, 
the effective anisotropy is suppressed by the exchange interaction
within a domain. In such materials the coercivity, $H_c$ and the grain size, $D$, are
related as \textit{i.e.,} $H_c \sim D^6$.
In contrast, when the grain size is
larger than the magnetic correlation length, the magnetization is controlled by
domain wall pinning at the grain boundaries and $H_c \sim 1/D$ \cite{goodenough-54,mager-52}.
This simple model has formed the basis of most theories on magnetic hardness in
nano-crystalline magnetic materials \cite{herzer-10}, and its predictions were
confirmed by experiments \cite{herzer-93,herzer-95,pfeifer-80}.

The relationship between grain size and hysteresis continues to be the
subject of numerous investigations in the field of micromagnetism
\cite{skomski-93,skomski-01,skomski-book,scherfl-00}. Most approaches have 
been based on minimization of Hamiltonians that contain terms for
anisotropy, exchange, magneto-elastic and demagnetizing energies
\cite{rave-97}. These Hamiltonians tend to be rather complex and can be solved
for minimum energy configurations only in simple geometries and under certain
simplifications, most of which do not adequately reflect the complex
microstructure of a material.

Magnetization has recently been coupled to phase field order parameters to
describe different crystal variants or phases. Magneto-phase field free
energies are then used to derive dynamical equations of microstructure
evolution in the presence of magnetic field in \cite{koyama-06, koyama-08,
zhang-05}. The advantage of these models is that the interface
between the crystalline phases emerges naturally in the dynamics of the field
equations. Furthermore, in this approach the free energy is developed in terms
of the coarse-grained order parameters in space and time
which makes it possible to reach the diffusive time scales
and the length scales necessary to study the microstructure evolution in the presence
of an external magnetic field.
The main drawback of traditional phase field approaches is that they
lack direct coupling to plastic and elastic effects emergent from the atomic
structure of crystalline phases. Information about the crystallographic
structure of material typically enters through effective parameters controlling
the anisotropy, exchange stiffness or the boundary conditions. Moreover, no
continuum phase field type model to date incorporates the elastic and kinetic
effects of topological defects.

This work introduces a new phase field crystal (PFC) model that captures the
basic physics of magneto-crystalline interactions for isotropic ferromagnetic
solids. The PFC approach \cite{elder-prl-02,elder-04} is a continuum method that has been
shown to capture the essential physics of atomic-scale
elastic and plastic effects that accompany diffusive phase transformations,
such as solidification, dislocation kinetics, solid state precipitation and
epitaxial growth \cite{elder-prb-07,provatas-07,
grant-epl-08,jaatinen-09,jaatinen-10,greenwood-10,greenwood-11,berry-12,fallah-12}. 
This work expands the approach by coupling the PFC density with
magnetization to generate a ferromagnetic solid below a density-dependent
Curie temperature. 
The equilibrium properties of the model are first examined, followed by a
single-mode derivation of the model's magnetostriction properties. Simple
dissipative dynamics are used to qualitatively illustrate coercivity to grain
and domain size. Finally, a relation for the coercivity versus low angle grain
boundary mis-orientation is derived. 


\section{Model description}
\label{sec:model}

We construct a dimensionless free energy that couples three dimensionless
fields, \textit{i.e.,} the number density $n(\mathbf x)$, the
magnetization vector $\mathbf m(\mathbf x)$ and the magnetic field $\mathbf B$, according to 
 
\begin{eqnarray}
 \frac{\Delta F}{k_BTV\bar{\rho}} \!\!  &=&  \!\! \mathop{\mathlarger{\int}} d \mathbf r \left \{   n 
\left (\Delta B+ B_s \left(1+ \nabla^2 \right)^2 \right) \frac{n}{2}- 
t\frac{n^3}{3}+v\frac{n^4}{4}   \right. \nonumber \\
& &+ \omega  \left [ \frac{W_0^2}{2} \left| \nabla
\mathbf m \right|^2+(r_c-\beta n^2) \frac{\left |\mathbf m \right |^2}{2}+ \gamma
\frac{|\mathbf m|^4}{4}  \right. \nonumber \\
& &\left. \left. -\frac{\alpha}{2} \left(\mathbf m \cdot \nabla n \right)^2 
-\mathbf m\cdot\mathbf B + \frac{\left |\mathbf B \right|^2}{2}  \right ] \right \}
\label{eq:chap7-mag-pfc-f-1}.
\end{eqnarray}
The dimensionless number density field is defined as $ n (\mathbf x) \equiv (\rho (\mathbf x) -\bar {\rho})/{\bar \rho}$, where 
$\bar{\rho}$ is a reference constant density, taken to be the density of the
liquid at coexistence \cite{nik-ken}.
 The parameter $\omega \equiv B_0^2/(\mu_0k_BT \bar{\rho})$ and 
$ B_0$ is a reference 
magnetic field. The parameter
$k_B$ is the Boltzmann constant and $V$ is the volume of the unit cell.

The first line in Eq.~\ref{eq:chap7-mag-pfc-f-1} is the usual phase field crystal (PFC) free energy
\cite{elder-prl-02, elder-04}, where $B_s$, $t$, $v$ and $\Delta B$ are dimensionless
parameters related to physical properties of the material. Using the classical density functional theory of freezing (CDFT) it has been shown that
$B_s$ is related to the bulk modulus of the solid \cite{elder-prb-07}.
In the same framework, the bulk parameters $t=1/2$ and $v=1/3$ are approximated by expanding the free energy
of the ideal gas about a reference density \cite{jaatinen-09,jaatinen-10}. 
In the PFC model the transition from liquid to solid occurs when $\Delta B$ changes.
Decreasing $\Delta B$ is equivalent to decreasing temperature or increasing the average density
of the system \cite{nik-ken}. Space has been rescaled in terms of the lattice constant $a=2 \pi/q_o$. 

The second line of the free energy approximates the magnetic free energy  through a Ginzburg-Landau (GL) expansion, 
which accounts for a ferromagnetic phase transition \cite{chaikin-lubensky}
through $(r_c-\beta n^2)$, a factor that implicitly depends on temperature
through $n$. This term defines the Curie temperatures. The parameters
$r_c$ and $\beta$ are chosen such that the Curie temperature lies below the
liquid/solid coexistence lines in the phase diagram \cite{herlach-98, herlach-01,
herlach-07}. The parameter $W_0$ sets the scale of the magnetic correlation
length (in units of the lattice constant). The parameter $\gamma$ is related to
the saturation magnetization and the magnetic susceptibility.
The term, $-\alpha/2 (\mathbf m \cdot \nabla n)^2$ is introduced in this work as the lowest
order coupling of magnetization to the gradient of the density,  giving rise to
magnetostriction, with $\alpha$ being the magnetostrictive coefficient, a very
small dimensionless quantity of order $\alpha \sim 10^{-5}$ for iron, nickel
and cobalt \cite{magnetostrictive}.

The last two terms in Eq.\,\ref{eq:chap7-mag-pfc-f-1} account for the the magnetic free
energy and are associated with the magnetic dipole interactions. 
The total local magnetic field is denoted by $\mathbf B=\mathbf B(\mathbf r)$, where $\mathbf B$ is scaled by
$B_0$. It is calculated by adding the induced magnetic field and the
external magnetic field, $\mathbf B(\mathbf r)=\mathbf B_{ind}(\mathbf
r)+\mathbf B_{ext}$. The induced magnetic field is a result of magnetization
current density, $\mathbf J_M$ and can be expressed in terms of the
magnetization, $\mathbf m$ according to the dimensionless Poisson's equation \cite{jackson}

\begin{equation}
\nabla^2 \mathbf A= - \nabla \times \mathbf m,
\label{eq:chap7-poisson}
\end{equation} 
where $\mathbf A$ is the dimensionless vector potential
and is related to the induced magnetic field via $\mathbf B_{ind}=\nabla \times
\mathbf A$.  
The value of $\omega$ can be chosen to match the measured value for a specific ferromagnetic element. Here, for simplicity we chose $\omega=1$.
In the subsequent sections we use the following model parameters in the free energy of
Eq.~(\ref{eq:chap7-mag-pfc-f-1}): $(B_s,t,v,\alpha,W_0,\beta,r_c,\omega,\gamma)=(0.98, 0.5, 1/3, 10^{-3},1, 4\times10^{-2}, 10^{-2},1, 1)$.

 \begin{figure}[htb]
\begin{center}
    \centerline{\includegraphics*[width=3.0in]{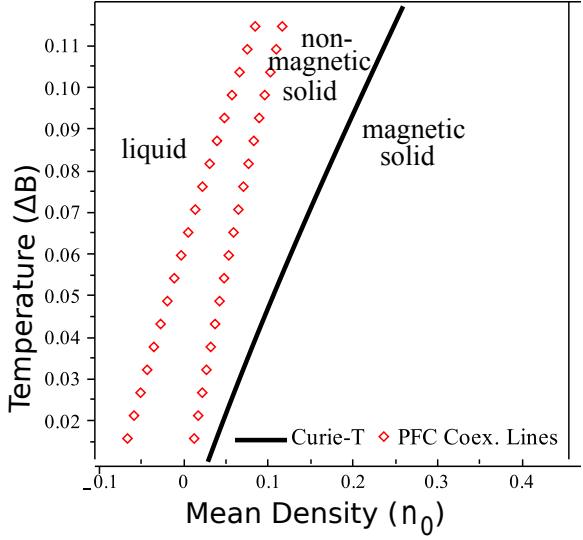}}
\caption[]{\linespread{0.75} Phase diagram of the Magnetic-PFC model. The dots
denote the liquid-solid coexistence lines and solid line denotes the Curie line
of ferromagnetic phase transition.} 
\label{fig:chap7-mpfc-pd}
\end{center}
\end{figure}
 
\section{Equilibrium properties and Magnetostriction}

In this section, an amplitude representation of
Eq.~\ref{eq:chap7-mag-pfc-f-1} is derived and then used to 
calculate the equilibrium phase diagram  and the magnetostriction
coefficients of the present model.
 
 \subsection{Phase Diagram}
To coarse-grain the free energy, we calculated the amplitude representation of
the free energy, by expanding the density in terms of the Fourier modes with
spatially-dependent complex amplitudes 
\begin{equation}
n=n_0+\sum_{j} \eta_j (\mathbf x,t) e^ {i\mathbf q_j\cdot \mathbf r} + C.C.
\label{eq:chap7-amp-exp-density}
\end{equation}
where $n_o$ represents the average density, $C.C.$ represents the complex
conjugate of the expansion, $\mathbf q_1=-\sqrt{3}/2\mathbf i-1/2\mathbf j $,
$\mathbf q_2=\mathbf j $ and $\mathbf q_3=\sqrt{3}/2\mathbf i-1/2\mathbf j$ for
a triangular two dimensional system and $\eta_i$ are complex amplitudes
corresponding to each density wave in the expansion.
Equation~(\ref{eq:chap7-amp-exp-density}) approximates the density to the
lowest order harmonics of the Fourier expansion, \textit{i.e.,} the single mode
approximation. We substitute this into the free energy (and omit the energy
associated with the magnetic dipole interactions in comparison to the exchange
interaction) and follow the
integration procedure of  \cite{goldenfeld-05, goldenfeld-06}, assuming that
the amplitudes, $\eta_j(\mathbf x,t)$, vary on length scale much larger than
the lattice constant.  In this assumption, we can simplify the calculations by
ignoring the spatial dependence of the amplitudes over a unit cell to obtain
\begin{equation}
\begin{split}
 &F _{\eta}  =   \int d \mathbf r \left\{   \bigg( \Delta B+B_s \bigg) \bigg (\frac{1}{2}n_0^2
+\sum_{j}|\eta_j|^2 \bigg) \right. \\
 &+B_s \bigg( -\sum_{j}|\eta_j|^2+\sum_{j}|\mathcal G_j\eta_j|^2 \bigg) \\
 &-t\left( \frac{1}{3}n_0^3+2n_0\sum_{j}|\eta_j|^2+2 \bigg( \prod _{j}\eta_j+\prod_{j} \eta_j^{\ast} \bigg) \right)  \\
 &+v \left[ \frac{1}{4}n_0^4+3n_0^2\sum_{j}|\eta_j|^2+\frac{3}{2}\sum_{j}|\eta_j|^4 \right.\\
 &+6(|\eta_1|^2|\eta_3|^2+|\eta_1|^2|\eta_2|^2+|\eta_2|^2|\eta_3|^2 )  \\
&+6n_0 \left. \bigg( \prod _{j}\eta_j+\prod_{j} \eta_j^{\ast}   \bigg) \right] -\omega \, \alpha \sum_{j} |\mathcal A_j \eta_j|^2  \\
&+\omega \left(  \frac{W_0^2}{2} |\nabla^2 \mathbf m|^2 
+  \bigg[ r_c-\beta \Big (n_0^2+2 \sum_{j}|\eta_j|^2 \Big) \bigg] \frac{|\mathbf m|^2}{2} \right.\\
& \left. \left. +\gamma \frac{|\mathbf m|^4}{4}  \right) \right\}, 
\end{split}
\label{eq:chap7-amplitude-expansion-model}
\end{equation}
where the operators $\mathcal G_j$ and $\mathcal A_j$ are defined as $\mathcal
G_j \equiv \left(\nabla^2+2i \mathbf q_j \cdot \nabla \right)$ and $\mathcal
A_j \equiv \mathbf m \cdot \left(\nabla + i \mathbf q_j  \right)$. 

Following Chan and Goldenfeld \cite{goldenfeld-09} and others \cite{elder-10}, we represent
the complex amplitude by $\eta_j=\phi e^{i \mathbf q_j \cdot \mathbf u}$, where
$\mathbf u (\mathbf x)$ denotes a local elastic displacement vector. 
Inserting it into Eq.\,\ref{eq:chap7-amplitude-expansion-model} we obtain
\begin{eqnarray}
& & F _{\phi,U_{ij}} = \int d \mathbf r \left \{  \Big( -\frac{t}{3}+ \frac{1}{2} (\Delta B+B_s) \Big) n_0^2 +\frac{v}{4}n_0^4 \right. \nonumber \\ 
& &+ \phi^2( 3\Delta B-6tn_0+9vn_0^2) +\phi^3 ( -4t+12vn_0) + \phi^4 \left(\frac{45}{2}v \right) \nonumber \\
& &+B_s \left[  \phi^2 \Big(  \frac{9}{2} ( U_{xx}^2+U_{yy}^2)+3U_{xx}U_{yy}+6U_{xy}^2   \Big)  \right] \nonumber \\
& &+\omega \left. \bigg[ -3\alpha{\phi}^2 (  U_{xx}m_x^2+U_{yy}m_y^2 + 2 U_{xy}m_xm_y ) \right. \nonumber \\
& &+\left(  \frac{W_0^2}{2} |\nabla \mathbf m|^2 +  \Big[ r_c-\beta(n_0^2+6\phi^2) \right.   \nonumber \\
& & \left. \left. \left. -3\alpha \phi^2 \Big]\frac{|\mathbf m|^2}{2}+\gamma \frac{|\mathbf m|^4}{4}  \right) \right ] \right \}, 
\label{eq:chap7-f-phi-u}
\end{eqnarray}
where $U_{ij}$ are the strain tensor elements and $\phi$ is the order
parameter. $\phi$ is non-zero in the solid phase and zero in the liquid phase. 
The strain tensor is defined as
\begin{equation}
U_{ij} \equiv \frac{1}{2}( u_{ij}+u_{ji}+\sum_{k} u_{ik} u_{jk} ), 
\end{equation}
with $u_{ab}\equiv \partial u_a/\partial x_b$. This form of the free energy
presents a continuum description of the magnetic material's properties which
has the same level of coarse-graining as those established in
\cite{landau-lifshitz-electrodynamics, atkin-elesticity}. The main difference
is that complex amplitudes now allow a description of grain boundaries and
dislocations.

To calculate the equilibrium states of the system, we minimize the free energy,
$F_{\phi, U_{ij}}$, with respect to the strain tensor elements, $U_{ij}$. The
resulting, minimized, free energy becomes
\begin{eqnarray}
&& F_{\phi} = \int d \mathbf r \left \{ \omega \left (\frac{W_0^2}{2} | \nabla\mathbf m |^2 +(- 3\alpha\phi^2-6\beta\phi^2 \right. \right. \nonumber \\
&& \left. \left. +r_c-\beta n_0^2 ) \frac{|\mathbf m|^2}{2} + \left(\gamma -\frac{9}{4}\frac{\phi^2\omega\alpha^2}{B_s} \right)\frac{|\mathbf m|^4}{4} \right ) \right. \nonumber \\
&& \left. +( 3\Delta B-6tn_0+9vn_0^2  )\phi^2+( 12vn_0-4t )\phi^3+\frac{45}{2} v \phi^4 \right. \nonumber \\
&&\left. +\left[ \frac{1}{2}(B_s+\Delta B ) -\frac{1}{3}t\right] n_0^2+\frac{1}{4}vn_0^4 \right \} 
\label{eq:chap7-mag-lg-model}.
\end{eqnarray}
This free energy has the form of the LG free energy containing a ferromagnetic
phase transition. The coefficients depend on the mean density of the system,
$n_0$, the crystalline order parameter ($\phi$) as well as the temperature, $\Delta B$.
The dependence of magnetic phase transition on the density is a manifestation
of the connection between microstructure and magnetic properties of the
material. 

The fact that the magnetic energy, Eq.\,\ref{eq:chap7-mag-lg-model}, does not
depend on the relative angle of the magnetization and the direction of $\nabla
n$, reveals that the model (Eq.\,\ref{eq:chap7-mag-pfc-f-1}) does not include
anisotropy. This is due to the fact that it contains the coupling term,
$\mathbf m \cdot \nabla n$, only up to the second order. To involve the
anisotropic effects higher order terms in $\mathbf m \cdot \nabla n$ are
required. Anisotropy will be examined in an upcoming paper.

To find the ferromagnetic transition, we minimize $F_{\phi}$ with respect to
the magnetization $|\mathbf m|$, obtaining
\begin{equation}
m_s = 0, \pm\sqrt{\frac{-\left(r_c-n_0^2\beta-6\phi^2\beta+3\phi^2\alpha\right)}
{\gamma-\frac{9}{4}\frac{\omega \, \alpha^2 \, \phi^2}{B_s}}}
\label{eq:chap7-m-min}
\end{equation}
where $\phi$ is the amplitude of the density field expansion in PFC model.
Minimizing the magnetic PFC free energy with respect to $\phi$ 
when $|\mathbf m|=0$ yields,
\begin{equation}
\phi_{min}=-\frac{n_0}{5}+\frac{1}{10}+\frac{1}{10}\sqrt{1+16n_0-16n_0^2-20\Delta B}
\label{eq:chap7-phi-min}
\end{equation}
Thus $\phi_{min}$  is a function of the mean density, $n_0$, and the temperature $\Delta B$.
To find the ferromagnetic transition line in the $\Delta B-n_0$ plane, we solve
\begin{equation}
r_c-n_0^2\beta-6\phi^2\beta+3\phi^2\alpha=0
\label{curie_exp}
\end{equation}
using $\phi=\phi_{min}$ in Eq.\,\ref{eq:chap7-phi-min}. The solution gives an
equation for the Curie line in terms of $\Delta B$ and average density $n_o$.
This line separates the phase diagram into non-magnetic and magnetic phases.

The values of $r_c$ and $\beta$ change the position of the Curie line in the
phase diagram. For a realistic model which is in accordance with the
experimental data \cite{herlach-98, herlach-01, herlach-07}, we choose them in
such a way that the curve lies below the PFC coexistence lines. This guarantees
that the ferromagnetic phase appear only in the solid. If we increase $r_c$ or
decrease $\beta$, the line will shift down.  The Curie line, together with the
solid-liquid coexistence lines of the model, are shown in
Fig.~\ref{fig:chap7-mpfc-pd}. The free energy is minimized by three phases:
liquid, non-magnetic solid and magnetic-solid. Above the Curie line $\mathbf
m=0$ and the usual PFC phase diagram reproduced.  

\subsection{Magentostriction}

The term $-\alpha/2(\mathbf m \cdot \nabla n)^2$ is a minimal coupling 
that induces magnetostriction. The strain energy is given by 
\begin{eqnarray}
&&{\mathcal F_{m-e}} = \mathop{\mathlarger{\int}} d\vec{r}  \left \{ B_s\left(9(U^2_{xx}+U^2_{yy})+6U_{xx}U_{yy}+12U^2_{xy}\right) \right. \nonumber \\
&& \left. -3 \omega \alpha \phi^2 \left( m^2_xU_{xx}+m^2_yU_{yy}+2m_xm_yU_{xy}  + \frac{|\mathbf m|^2}{2} \right) \right \}. \nonumber \\ \nonumber \\
\label{Fme}
\end{eqnarray}
Minimizing the resultant free energy with respect to the strain tensor
elements yields the relations for magnetically induced strain tensor as
\begin{eqnarray}
\label{eq:chap7-mag-elas-strain-4}
U_{xx}^{min}&=&\frac{\alpha \, \omega \, m^2}{8B_s}(2\cos2\theta+1)\\
\label{eq:chap7-mag-elas-strain-5}
U_{yy}^{min}&=&-\frac{\alpha \, \omega \, m^2}{8B_s}(2\cos2\theta-1)\\
\label{eq:chap7-mag-elas-strain-6}
U_{xy}^{min}&=&\frac{\alpha \, \omega \, m^2}{4B_s}\sin2\theta,
\end{eqnarray}
where we have assumed that the magnetization vector has an angle $\theta$
with respect to the $x$-axis.

 \begin{figure}[htb]
\begin{center}
    \centerline{\includegraphics*[width=3.5in,height=4in]{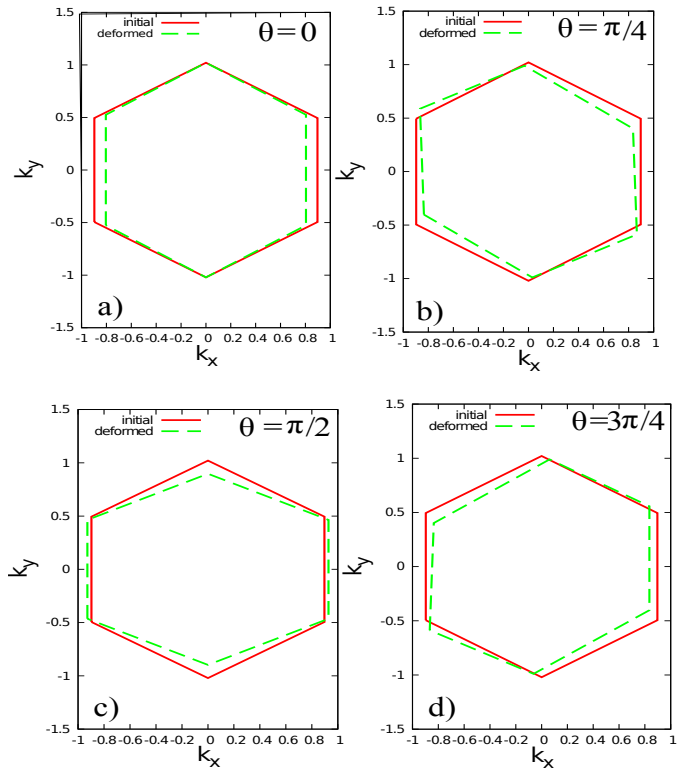}}
\caption[]{\linespread{0.75} The structure factor of the system of size $256
\times 256$ at coexistence. We first allow the system to equilibrate in the
absence of the magnetic field and then apply a magnetic field of $|\mathbf B| =
9.5$ at the angle of $\theta$ with respect to the $x$-axis. The red hexagons
show the structure factor of the initial configuration and the the green deformed
hexagons in $(a)-(d)$ show the magnetically induced deformation of the
hexagonal lattice at different magnetic field directions.
}
\label{fig:sf}
\end{center}
\end{figure}

	To examine the influence of magnetostriction on the crystalline states 
numerical simulations were conducted. In this work we were mainly interested 
in the equilibrated states and consequently used the simplest dissipative dynamics 
for both fields, such that conserved (model $B$ \cite{hohenberg-77}) 
dynamics were used for $n(\mathbf x)$ and non-conserved (model $A$ \cite{hohenberg-77}) 
dynamics for the magnetization, i.e.,  
\begin{eqnarray}
\label{chap7-eqs-of-motion-1}
\frac{\partial n(\mathbf r,t)}{\partial t}&=&
\nabla^2  \frac{\delta F(n,\mathbf m)}{\delta n}  \\
\label{chap7-eqs-of-motion-2-3}
 \frac{\partial m_i(\mathbf r,t)}{\partial t}&=&
 -\frac{\delta F(n,\mathbf m)}{\delta m_i},
\end{eqnarray}
where $i=x,y$ for the magnetization in $x$ and $y$ directions, respectively.
These equations were solved in a system with periodic boundary conditions 
using Euler's method for the time derivative, finite difference methods for 
the spatial gradients \cite{patra-05} and a Fourier transform algorithm to solve the Poisson 
equation for the vector potential $\mathbf A$. 

Since the magnetically induced strain is much smaller than the elastic strain
that can developed in a confined solid, this simulation was performed in a
system of coexisting liquid and solid phases.  In this instance the solid 
can deform freely when an external magnetic
field is applied. 
The system parameters are as follows: $(B_s,t,v,\alpha,W_0,\beta,r_c,\omega,\gamma)=(0.98, 0.5, 1/3, 10^{-3},1, 4\times10^{-2}, 10^{-2},1, 1)$.
We chose these values so that we have a wide a coexistence region for as large
a value of $\alpha$ as possible, to make the magnetostriction large enough to be
measurable. The coexistence
region lies above the Curie line in the phase diagram and is paramagnetic. When
an external magnetic field is applied, it will break the symmetry of the
magnetic free energy, and the net magnetization of the system becomes 
non-zero, aligning with the external magnetic field.

An external magnetic field of $|\mathbf B|=9.5$ was applied in different angles
with respect to $x$-axis. Figure \ref{fig:sf} shows the structure of the
hexagonal solid phase for angles $\theta=0, \pi/4, \pi/2, 3\pi/2$. In these
simulations $\alpha=-0.7$ is negative which means that the sample shrinks in
the direction of the applied magnetic field. The two hexagons in the these
figures represent the initial lattice points (red) and the 
deformed lattice points (green) after the magnetic field is applied.  
Figure \ref{fig:sf} clearly shows the magnetostriction effect.

\section{Effect of Grain Boundaries on Coercivity}

	In this section we examine the influence of grain boundaries and size on 
the magnetic coercivity.  As a baseline, we first study 
the mean field coercivity of a single crystal in subsection \ref{sec:cm} 
and then numerically examine the influence of grain boundary misorientation and grain size 
in subsections \ref{sec:co} and \ref{sec:cg} respectively.

\subsection{ Mean field coercivity of a single crystal} 
\label{sec:cm}

\begin{figure}[htb]
\begin{center}
    \centerline{\includegraphics*[width=2.5in ,height=2in]{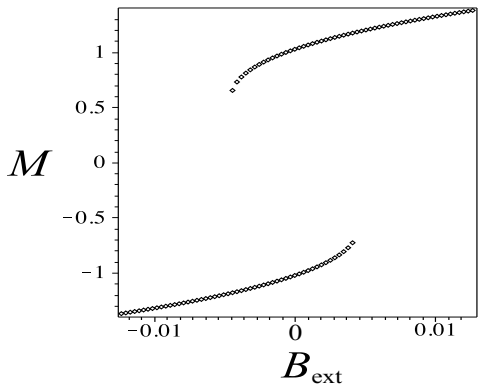}}
\caption[]{\linespread{0.75} The vertical axis represents the magnetization $M=m/m_s$
and the dotted lines represent the two roots of the magnetic free energy
$f_{\phi}$ given in Eq.\,\ref{eq:chap7-mag-lg-model}, corresponding to positive and negative magnetization values that
minimize the free energy. When an external magnetic field is applied, the spins
tend to flip to align with the magnetic field. If the magnetic field is large
enough $(|\mathbf B|>B_a=B_a=4.7 \times 10^{-3})$ one of the minima disappears. The
value of $B_a$ gives an estimation for the coercivity of the model with parameters 
$(B_s,t,v,\alpha,W_0,\beta,r_c,\omega,\gamma)=(0.98, 0.5, 1/3, 10^{-3},1, 4\times10^{-2}, 10^{-2},1, 1)$.}
\label{fig:chap8-roots}
\end{center}
\end{figure}

	The coercivity of a single crystal in the mean field limit can 
be obtained by determining when the local minima of the free energy with 
respect to $m$ disappears as a function of applied magnetic field $\mathbf B$.  The 
value of $\mathbf B$ at which this just occurs is the mean field coercivity. If 
thermal fluctuations were included a small coercivity would be obtained that 
would depend on how rapidly the applied field changes.  To obtain the 
mean field coercivity we first solve 
$d (f_{\phi}-\mathbf m \cdot \mathbf B)/dm=0$, where $f_{\phi}$ is the non-gradient 
part of the magnetic free energy, Eq.\,\ref{eq:chap7-mag-lg-model}, to obtain the minima 
of the free energy for $\mathbf B = B\, \hat{x}$.
These two minima are shown in Fig.\,\ref{fig:chap8-roots}
for different values of external magnetic field.  For $B<0$ ($B>0$) the negative (positive) 
branch has the lower free energy and positive (negative) branch is metastable.  The 
metastable branch disappears at the mean field coercivity. As shown in 
Fig.\,\ref{fig:chap8-roots} the coercivity is equal to $H_a=4.7 \times 10^{-3}$. 
The parameters of the free energy used in this calculation and in all of the simulations
of the following sections are as stated in Section \ref{sec:model}.

 \begin{figure}[htb]
\begin{center}
    \centerline{\includegraphics*[width=3in]{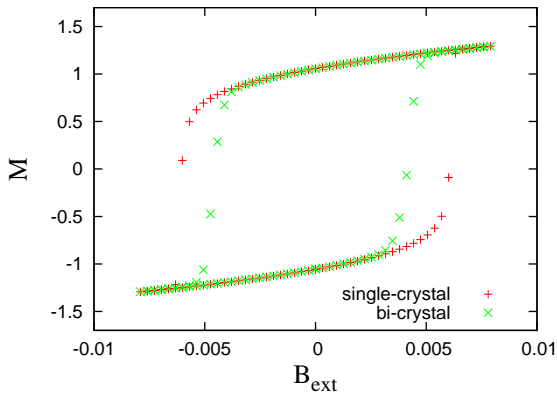}}
\caption[]{\linespread{0.75}  Hysteresis curves for single crystal and
bi-crystalline systems. The two-grain system is produced by placing two
hexagonal crystal lattices that are rotated symmetrically with respect to each other.
To produce the perfect fit inside the boxes, the lattice spacing was chosen to be
$dx=28 \pi/(64 \sqrt{3})$ and the time step $dt=10^{-3}$. It
can be seen that the coercivity of a single crystal is larger than that
of a bi-crystal.}
\label{fig:chap8-hys-comp-g12-paper}
\end{center}
\end{figure}

	To examine the hysteresis behavior and magnetic coercivity, numerical 
simulations were conducted
by solving the dynamical equations of motion, \textit{i.e., }
Eqs.\,\ref{chap7-eqs-of-motion-1} and \ref{chap7-eqs-of-motion-2-3}
using periodic boundary conditions.
The external magnetic field was increased linearly 
from zero to a maximum of $\mathbf B=(0,0.008)$ (this value was estimated using
Fig.\,\ref{fig:chap8-roots}) and was then decreased at the same rate to complete a cycle
in total time steps of $4\times10^5$. The field was applied to a single crystal system and 
a bi-crystal system. The bi-crystal system is prepared by
locating two symmetrically tilted grains, with a tilt angle of
$\theta=21.78$ degrees. With this choice for the mismatch angle,
the simulation box of size $256 \times 256$ accommodates two perfect grains and
the boundary effects does not disturb the energy of
the grain boundary.

We calculated the mean magnetization of the two systems as a
function of the applied magnetic field. The result is shown in
Fig.\,\ref{fig:chap8-hys-comp-g12-paper}. It can be seen that 
the coercivity is larger for a single crystal comparing to the bi-crystal system.
\textit{i.e.,} it takes a smaller external magnetic field to remove the local minimum of the free energy in the two-grain system, comparing
to the single crystal. This implies that the existence of a grain boundary facilitates the formation of the magnetic domains. This is a quite interesting result which can be also explained by consulting Eq.\,\ref{curie_exp}. Namely, since in the grain boundary region where $\phi=0$, the temperature at which the ferromagnetism appears is smaller than that in the bulk. Also, the grains are the sites of the system from which the magnetic domains start to from and grow. 
This affects the hysteresis curve and the coercivity of the system and the two-grain system has a smaller coercivity in comparison with the single crystal system.

\subsection{Coercivity and grain boundary misorientation }
\label{sec:co}

	In this section the influence of the grain boundary 
orientation on the coercivity is examined.  Low angle boundaries are 
characterized by a line of dislocations separated by a distance that is 
inversely proportional to the misorientation 
angle \cite{read-shockley}, while large angle boundaries are essentially 
a continuous region of disorder.  Typically dislocation cores lower the 
energy barrier for nucleation of magnetization to the lower energy state \cite{goodenough-54}.
Thus the nature of the grain boundary is expected to play a key 
role in determining the coercivity.
\begin{figure}[htb]
\begin{center}
    \centerline{\includegraphics*[width=2.5in]{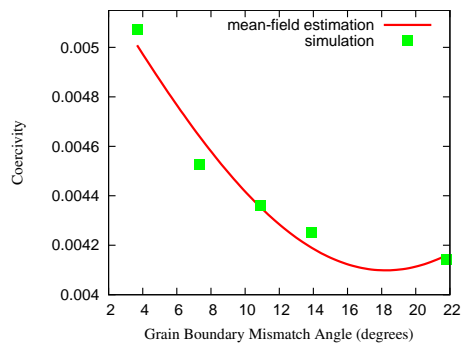}}
\caption[]{\linespread{0.75}  The functional form
$(a_1-a_2 \sin(a_3\theta))^{3/2}$ fitted to the simulation data of the coercivity
vs. the grain boundary angle for a system of $L=256$. The fit parameters are
$a_1=0.005$, $a_2=0.001$, and $a_3=0.08$. The coercivity
decreases as we increase the grain boundary angle as predicted by the
mean-field estimation. }
\label{fig:coercivity-angle-fit-sin}
\end{center}
\end{figure}

	 To examine this phenomena we performed simulations 
by preparing a bi-crystal inside with grains that have symmetric tilt angles.
The tilt angles are chosen in such a way that a perfect crystal
fits inside the simulation box to prevent
any change in the energy of the system regarding the boundary effects.
The angles that allow perfect crystal fit inside a box of
size $256 \times 256$ are $\theta=3.67, 7.31, 
10.89, 21.78$ degrees. Simulations are conducted 
similar the previous section.
 
        Simulations results of the coercivity for different grain boundary angles (green pluses) are
presented in Fig.\,\ref{fig:coercivity-angle-fit-sin}.
According to these data as the grain boundary mismatch angle increases, 
the coercivity of the system decreases. This is in accordance with the mean field arguments
of Section \ref{sec:cm} and the corresponding results of Fig.~\ref{fig:chap8-hys-comp-g12-paper}.
A single-crystal system, being the limiting case of the bi-crystal system as $\theta \rightarrow 0$,
has the largest coercivity comparing with the bi-crystal systems.
The existence of the grain boundary suppresses the the barrier to be overcome
and a magnetic domain of opposite direction initiate from the grain boundary region.

     These results and the arguments of Section \ref{sec:cm} suggest that
to approximate the coercivity in the presence of a grain boundary at a mean-field level, 
we replace the square of the phase amplitude $\phi^2$ with $\phi^2-NA/L$ in the magnetic
free energy of the system, Eq.\,\ref{eq:chap7-mag-lg-model}, where $N$ is the
number of dislocations per unit length of the grain boundary, $A$ is the unit
area influenced by a dislocation and $L$ is the size of the system.  Minimizing
the free energy with respect to the magnetization, in the presence of an
external magnetic field, $\mathbf H$, and noting that the number of
dislocations in a grain boundary is related to the mismatch angle as $N \sim
\sin (\theta/2)$ \cite{read-shockley}, we obtain

\begin{equation}
\begin{split}
H_c=& \sqrt{16B_s/(243 \, \omega \, \alpha^2\phi^2+108 \,\gamma \, B_s)} \\
 & \times [ (3\alpha+6\beta)(\phi^2-(cA/L)\sin(\theta/2))+r_c-\beta n_0^2] ^{3/2}
\end{split}
\label{H_ctheory}
\end{equation} 
for the coercivity as a function of the mismatch angle $\theta$. In this
equation $c \equiv 2 \cos \delta/a$, where $\delta$ is the angle of the grain
boundary with respect to the $x$ axis and $a$ is the lattice constant
\cite{read-shockley}. 

Figure~\ref{fig:coercivity-angle-fit-sin} shows a good fit of the simulation data with the mean-field relation in Eq.~\ref{H_ctheory}, confirming that the model is capturing the role of
grain boundary in magnetization process and coercivity.

\subsection{ Coercivity and grain size} 
\label{sec:cg}

\begin{figure}[htb]
\begin{center}
    \centerline{\includegraphics*[width=3in]{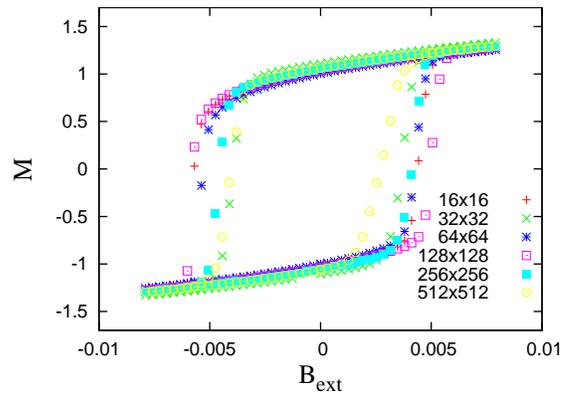}}
\caption[]{\linespread{0.75}  Hysteresis curves for two-grain systems for
different grain sizes. }
\label{fig:chap8-hys-size-w1-1}
\end{center}
\end{figure}

	To examine the hysteresis behavior and magnetic coercivity numerical 
simulations were conducted by applying an external
field to bi-crystal systems of different sizes.  
The mean magnetization of the system as a
function of the applied magnetic field for this situation is presented in
Fig.\,\ref{fig:chap8-hys-size-w1-1}. It can be seen that the width of the
hysteresis curves change with grain size. The mean magnetization was 
calculated after letting the system evolve to $4 \times 10^5$ simulation 
time steps at each value of the $\mathbf B$ field.

\begin{figure}[htb]
\begin{center}
    \centerline{\includegraphics*[width=2.5in]{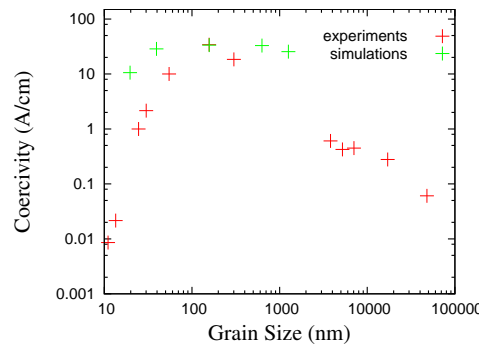}}
\caption[]{\linespread{0.75}  Simulation coercivity data as a function of
system size compared with the experimental data \cite{herzer-90,pfeifer-80}. To
make the comparison more clear, the coercivity values from the simulation data
are scaled by a factor of $5.5\times10^3$ to better compare the peak positions of both
data sets.}
\label{fig:coer-size-comp}
\end{center}
\end{figure}

The coercivity extracted from the hysteresis curves is plotted in
Fig.\,\ref{fig:coer-size-comp} as a function of grain size together with the
rescaled experimental data reported in \cite{herzer-93, herzer-97}. 
The data suggest that the grain size influences the coercivity even in the absence of anisotropy. However, we obtain 
qualitatively different results from experiments in the limit when the grain size is much larger than the magnetic correlation length.

In our simulations, as the grain size increases, the coercivity also increases, until it reaches to a maximum value. 
In this regime, the grain size is smaller than the magnetic correlation length.
As grain size increases in a fixed system size, the ratio of  grain boundary area to grain crystalline bulk area
 decreases. This implies less area available for
nucleation of reverse magnetization and the coercivity increases. Also, since anisotropy effect is not included in this model we do not expect quantitative agreement of simulation data points with experimental results
for large grain size regime.

 It should be noted that in this work, the comparisons in
Figs.~\ref{fig:chap8-hys-size-w1-1} and \ref{fig:coer-size-comp} are
qualitative. In the absence of magneto-crystalline anisotropy, it is expected
that the corecivity will in fact tend to zero after a long enough equilibration
time in the dynamics. Nevertheless,  the hysteresis results reported here
- essentially a meta-stable kinetic effect - point to the robustness of the model
to capture some relevant features.  
A more quantitative examination of
coercivity that includes anisotropy will be reported in an upcoming paper.  

\section{Discussion and conclusions}
The Landau-Lifshitz-Gilbert (LLG) equation is perhaps the most 
common approach to modelling the dynamics of the magnetization vector. The LLG 
approach and its many variants has been used to study domain walls, effect of 
disorder and it has been even connected to density functional theory 
computations \cite{iwasaki2013,chimata,fal2013}, for a review see, e.g., Kurzik and 
Prohl~\cite{kruzik2006}. In LLG, the dynamic equations are solved in a lattice 
but there is no direct connection with the underlaying atomic structure. That 
is different in our approach. Here, we couple of the density field representing 
the atoms to the magnetic moments and our approach includes both the Curie 
and melting/solidification in its phase diagram. 
The LLG dynamics includes the precessional motion of magnetization which is important on short time scales.
By Using Model A dynamics we are limiting our attention to diffusive time scales which are relevant
to phase transformations. It is noted that our approach 
(although beyond the scope of this work), can be extended to cases such as 
magnetization of binary alloys and inclusion of impurities and pinning.

In summary, we extended the PFC formalism \cite{elder-04} to include isotropic
magneto-crystalline interactions for the study of ferromagnetic solids. 
The advantage of this model over other phase field models of magneto-elastic 
systems is that it naturally incorporates atomistic features such as 
dislocations and grain boundaries.  As a result many macroscopic properties, 
such as coercivity, can be linked to details of grain boundaries 
and polycrystalline structures.  While this initial studies was for a one 
component two dimensional system  that was magnetically isotropic it 
is straightforward to extend the model to include anisotropy, binary alloys and
three dimensions.

\section{Acknowledgment}
This work was supported by the Natural Science and Engineering Research
Council of Canada (NSERC). We are also grateful to Compute Canada (SHARCNET and the McGill HPC Centre) 
for providing the computing resources. KRE acknowledges support from  NSF under Grant No. DMR-0906676.


\begin{thebibliography}{55}
\expandafter\ifx\csname natexlab\endcsname\relax\def\natexlab#1{#1}\fi
\expandafter\ifx\csname bibnamefont\endcsname\relax
  \def\bibnamefont#1{#1}\fi
\expandafter\ifx\csname bibfnamefont\endcsname\relax
  \def\bibfnamefont#1{#1}\fi
\expandafter\ifx\csname citenamefont\endcsname\relax
  \def\citenamefont#1{#1}\fi
\expandafter\ifx\csname url\endcsname\relax
  \def\url#1{\texttt{#1}}\fi
\expandafter\ifx\csname urlprefix\endcsname\relax\def\urlprefix{URL }\fi
\providecommand{\bibinfo}[2]{#2}
\providecommand{\eprint}[2][]{\url{#2}}

\bibitem[{\citenamefont{McHenry et~al.}(1999)\citenamefont{McHenry, Willard,
  and Laughlin}}]{mchenry-99}
\bibinfo{author}{\bibfnamefont{M.~E.} \bibnamefont{McHenry}},
  \bibinfo{author}{\bibfnamefont{M.~A.} \bibnamefont{Willard}},
  \bibnamefont{and} \bibinfo{author}{\bibfnamefont{D.~E.}
  \bibnamefont{Laughlin}}, \bibinfo{journal}{Prog. Mater. Sci.}
  \textbf{\bibinfo{volume}{44}}, \bibinfo{pages}{291} (\bibinfo{year}{1999}).

\bibitem[{\citenamefont{McHenry and Laughlin}(2000)}]{mchenry-00}
\bibinfo{author}{\bibfnamefont{M.~E.} \bibnamefont{McHenry}} \bibnamefont{and}
  \bibinfo{author}{\bibfnamefont{D.~E.} \bibnamefont{Laughlin}},
  \bibinfo{journal}{Acta Mater.} \textbf{\bibinfo{volume}{48}},
  \bibinfo{pages}{223} (\bibinfo{year}{2000}).

\bibitem[{\citenamefont{Provatas et~al.}(2007)\citenamefont{Provatas, Dantzig,
  Athreya, Chan, Stefanovic, Goldenfeld, and Elder}}]{provatas-07}
\bibinfo{author}{\bibfnamefont{N.}~\bibnamefont{Provatas}},
  \bibinfo{author}{\bibfnamefont{J.~A.} \bibnamefont{Dantzig}},
  \bibinfo{author}{\bibfnamefont{B.}~\bibnamefont{Athreya}},
  \bibinfo{author}{\bibfnamefont{P.}~\bibnamefont{Chan}},
  \bibinfo{author}{\bibfnamefont{P.}~\bibnamefont{Stefanovic}},
  \bibinfo{author}{\bibfnamefont{N.}~\bibnamefont{Goldenfeld}},
  \bibnamefont{and} \bibinfo{author}{\bibfnamefont{K.~R.} \bibnamefont{Elder}},
  \bibinfo{journal}{JOM} \textbf{\bibinfo{volume}{59}}, \bibinfo{pages}{83}
  (\bibinfo{year}{2007}).

\bibitem[{\citenamefont{White}(1985)}]{white-85}
\bibinfo{author}{\bibfnamefont{R.~M.} \bibnamefont{White}},
  \bibinfo{journal}{Science} \textbf{\bibinfo{volume}{229}},
  \bibinfo{pages}{11} (\bibinfo{year}{1985}).

\bibitem[{\citenamefont{Fukunaga and Inoue}(1992)}]{fukunaga-92}
\bibinfo{author}{\bibfnamefont{H.}~\bibnamefont{Fukunaga}} \bibnamefont{and}
  \bibinfo{author}{\bibfnamefont{H.}~\bibnamefont{Inoue}},
  \bibinfo{journal}{Jpn. J. Appl. Phys.} \textbf{\bibinfo{volume}{31}},
  \bibinfo{pages}{1347} (\bibinfo{year}{1992}).

\bibitem[{\citenamefont{Rave and Ramstok}(1997)}]{rave-97}
\bibinfo{author}{\bibfnamefont{W.}~\bibnamefont{Rave}} \bibnamefont{and}
  \bibinfo{author}{\bibfnamefont{K.}~\bibnamefont{Ramstok}},
  \bibinfo{journal}{J. Magn. Magn. Mater.} \textbf{\bibinfo{volume}{171}},
  \bibinfo{pages}{69} (\bibinfo{year}{1997}).

\bibitem[{\citenamefont{Skomski et~al.}(2001)\citenamefont{Skomski, Zeng, and
  Sellmyer}}]{skomski-01}
\bibinfo{author}{\bibfnamefont{R.}~\bibnamefont{Skomski}},
  \bibinfo{author}{\bibfnamefont{H.}~\bibnamefont{Zeng}}, \bibnamefont{and}
  \bibinfo{author}{\bibfnamefont{D.~J.} \bibnamefont{Sellmyer}},
  \bibinfo{journal}{IEEE Trans. Magn.} \textbf{\bibinfo{volume}{37}},
  \bibinfo{pages}{2549} (\bibinfo{year}{2001}).

\bibitem[{\citenamefont{Hernando and Gonzalez}(2000)}]{hernando-00}
\bibinfo{author}{\bibfnamefont{A.}~\bibnamefont{Hernando}} \bibnamefont{and}
  \bibinfo{author}{\bibfnamefont{J.~M.} \bibnamefont{Gonzalez}},
  \bibinfo{journal}{Hyperfine Interact.} \textbf{\bibinfo{volume}{130}},
  \bibinfo{pages}{221} (\bibinfo{year}{2000}).

\bibitem[{\citenamefont{Herzer}(1997)}]{herzer-97}
\bibinfo{author}{\bibfnamefont{G.}~\bibnamefont{Herzer}}, in
  \emph{\bibinfo{booktitle}{Handbook of Magnetic Materials}}, edited by
  \bibinfo{editor}{\bibfnamefont{K.~H.~J.} \bibnamefont{Buschow}}
  (\bibinfo{publisher}{Elsevier Science B.V.}, \bibinfo{year}{1997}),
  vol.~\bibinfo{volume}{10}, chap.~\bibinfo{chapter}{3}.

\bibitem[{\citenamefont{Fidler and Scherfl}(2000)}]{scherfl-00}
\bibinfo{author}{\bibfnamefont{J.}~\bibnamefont{Fidler}} \bibnamefont{and}
  \bibinfo{author}{\bibfnamefont{T.}~\bibnamefont{Scherfl}},
  \bibinfo{journal}{J. Phys. D: Appl. Phys.} \textbf{\bibinfo{volume}{33}},
  \bibinfo{pages}{R135} (\bibinfo{year}{2000}).

\bibitem[{\citenamefont{Harris et~al.}(1973)\citenamefont{Harris, Plischke, and
  Zuckermann}}]{zuckermann-73}
\bibinfo{author}{\bibfnamefont{R.}~\bibnamefont{Harris}},
  \bibinfo{author}{\bibfnamefont{M.}~\bibnamefont{Plischke}}, \bibnamefont{and}
  \bibinfo{author}{\bibfnamefont{M.~J.} \bibnamefont{Zuckermann}},
  \bibinfo{journal}{Phys. Rev. Lett.} \textbf{\bibinfo{volume}{31}},
  \bibinfo{pages}{160} (\bibinfo{year}{1973}).

\bibitem[{\citenamefont{Alben et~al.}(1978)\citenamefont{Alben, Becker, and
  Chi}}]{alben-becker-chi-78}
\bibinfo{author}{\bibfnamefont{R.}~\bibnamefont{Alben}},
  \bibinfo{author}{\bibfnamefont{J.~J.} \bibnamefont{Becker}},
  \bibnamefont{and} \bibinfo{author}{\bibfnamefont{M.~C.~.} \bibnamefont{Chi}},
  \bibinfo{journal}{J. Appl. Phys.} \textbf{\bibinfo{volume}{49}},
  \bibinfo{pages}{1653} (\bibinfo{year}{1978}).

\bibitem[{\citenamefont{Chi and Alben}(1977)}]{chi-alben-77}
\bibinfo{author}{\bibfnamefont{M.~C.} \bibnamefont{Chi}} \bibnamefont{and}
  \bibinfo{author}{\bibfnamefont{R.}~\bibnamefont{Alben}}, \bibinfo{journal}{J.
  Appl. Phys.} \textbf{\bibinfo{volume}{48}}, \bibinfo{pages}{1653}
  (\bibinfo{year}{1977}).

\bibitem[{\citenamefont{Goodenough}(1954)}]{goodenough-54}
\bibinfo{author}{\bibfnamefont{J.~B.} \bibnamefont{Goodenough}},
  \bibinfo{journal}{Phys. Rev.} \textbf{\bibinfo{volume}{95}},
  \bibinfo{pages}{917} (\bibinfo{year}{1954}).

\bibitem[{\citenamefont{Mager}(1952)}]{mager-52}
\bibinfo{author}{\bibfnamefont{A.}~\bibnamefont{Mager}}, \bibinfo{journal}{Ann.
  Physik} \textbf{\bibinfo{volume}{6. {Folge. Band 11}}}, \bibinfo{pages}{15}
  (\bibinfo{year}{1952}).

\bibitem[{\citenamefont{Flohrer and Herzer}(2010)}]{herzer-10}
\bibinfo{author}{\bibfnamefont{S.}~\bibnamefont{Flohrer}} \bibnamefont{and}
  \bibinfo{author}{\bibfnamefont{G.}~\bibnamefont{Herzer}},
  \bibinfo{journal}{J. Magn. Magn. Mater.} \textbf{\bibinfo{volume}{322}},
  \bibinfo{pages}{1511} (\bibinfo{year}{2010}).

\bibitem[{\citenamefont{Herzer}(1993)}]{herzer-93}
\bibinfo{author}{\bibfnamefont{G.}~\bibnamefont{Herzer}},
  \bibinfo{journal}{Phys. Scr.} \textbf{\bibinfo{volume}{T49A}},
  \bibinfo{pages}{307} (\bibinfo{year}{1993}).

\bibitem[{\citenamefont{Herzer}(1995)}]{herzer-95}
\bibinfo{author}{\bibfnamefont{G.}~\bibnamefont{Herzer}},
  \bibinfo{journal}{Scripta Metall Mater} \textbf{\bibinfo{volume}{33}},
  \bibinfo{pages}{1741} (\bibinfo{year}{1995}).

\bibitem[{\citenamefont{Pfeifer and Redeloff}(1980)}]{pfeifer-80}
\bibinfo{author}{\bibfnamefont{F.}~\bibnamefont{Pfeifer}} \bibnamefont{and}
  \bibinfo{author}{\bibfnamefont{C.}~\bibnamefont{Redeloff}},
  \bibinfo{journal}{J. Magn. Magn. Mater} \textbf{\bibinfo{volume}{19}},
  \bibinfo{pages}{190} (\bibinfo{year}{1980}).

\bibitem[{\citenamefont{Skomski and Coey}(1993)}]{skomski-93}
\bibinfo{author}{\bibfnamefont{R.}~\bibnamefont{Skomski}} \bibnamefont{and}
  \bibinfo{author}{\bibfnamefont{M.~D.} \bibnamefont{Coey}},
  \bibinfo{journal}{Phys. Rev. B} \textbf{\bibinfo{volume}{48}},
  \bibinfo{pages}{15812} (\bibinfo{year}{1993}).

\bibitem[{\citenamefont{Skomski}(2008)}]{skomski-book}
\bibinfo{author}{\bibfnamefont{R.}~\bibnamefont{Skomski}},
  \emph{\bibinfo{title}{Simple Models of Magnetism}}
  (\bibinfo{publisher}{Oxford University Press, New York},
  \bibinfo{year}{2008}).

\bibitem[{\citenamefont{Koyama and Onodera}(2006)}]{koyama-06}
\bibinfo{author}{\bibfnamefont{T.}~\bibnamefont{Koyama}} \bibnamefont{and}
  \bibinfo{author}{\bibfnamefont{H.}~\bibnamefont{Onodera}},
  \bibinfo{journal}{J. Phase Equilib. Diff.} \textbf{\bibinfo{volume}{27}},
  \bibinfo{pages}{22} (\bibinfo{year}{2006}).

\bibitem[{\citenamefont{Koyama}(2008)}]{koyama-08}
\bibinfo{author}{\bibfnamefont{T.}~\bibnamefont{Koyama}},
  \bibinfo{journal}{Sci. Technol. Adv. Mater.} \textbf{\bibinfo{volume}{9}},
  \bibinfo{pages}{013006} (\bibinfo{year}{2008}).

\bibitem[{\citenamefont{Zhang and Chen}(2005)}]{zhang-05}
\bibinfo{author}{\bibfnamefont{J.~X.} \bibnamefont{Zhang}} \bibnamefont{and}
  \bibinfo{author}{\bibfnamefont{L.~Q.} \bibnamefont{Chen}},
  \bibinfo{journal}{Acta Materialia} \textbf{\bibinfo{volume}{2845}},
  \bibinfo{pages}{53} (\bibinfo{year}{2005}).

\bibitem[{\citenamefont{Elder et~al.}(2002)\citenamefont{Elder, Katakowski,
  Haataja, and Grant}}]{elder-prl-02}
\bibinfo{author}{\bibfnamefont{K.~R.} \bibnamefont{Elder}},
  \bibinfo{author}{\bibfnamefont{M.}~\bibnamefont{Katakowski}},
  \bibinfo{author}{\bibfnamefont{M.}~\bibnamefont{Haataja}}, \bibnamefont{and}
  \bibinfo{author}{\bibfnamefont{M.}~\bibnamefont{Grant}},
  \bibinfo{journal}{Phys. Rev. Lett.} \textbf{\bibinfo{volume}{88}},
  \bibinfo{pages}{245701} (\bibinfo{year}{2002}).

\bibitem[{\citenamefont{Elder and Grant}(2004)}]{elder-04}
\bibinfo{author}{\bibfnamefont{K.~R.} \bibnamefont{Elder}} \bibnamefont{and}
  \bibinfo{author}{\bibfnamefont{M.}~\bibnamefont{Grant}},
  \bibinfo{journal}{Phys. Rev. E} \textbf{\bibinfo{volume}{70}},
  \bibinfo{pages}{051605} (\bibinfo{year}{2004}).

\bibitem[{\citenamefont{Elder et~al.}(2007)\citenamefont{Elder, Provatas,
  Berry, Stefanovic, and Grant}}]{elder-prb-07}
\bibinfo{author}{\bibfnamefont{K.~R.} \bibnamefont{Elder}},
  \bibinfo{author}{\bibfnamefont{N.}~\bibnamefont{Provatas}},
  \bibinfo{author}{\bibfnamefont{J.}~\bibnamefont{Berry}},
  \bibinfo{author}{\bibfnamefont{P.}~\bibnamefont{Stefanovic}},
  \bibnamefont{and} \bibinfo{author}{\bibfnamefont{M.}~\bibnamefont{Grant}},
  \bibinfo{journal}{Phys. Rev. B} \textbf{\bibinfo{volume}{75}},
  \bibinfo{pages}{064107} (\bibinfo{year}{2007}).

\bibitem[{\citenamefont{Tupper and Grant}(2008)}]{grant-epl-08}
\bibinfo{author}{\bibfnamefont{P.~F.} \bibnamefont{Tupper}} \bibnamefont{and}
  \bibinfo{author}{\bibfnamefont{M.}~\bibnamefont{Grant}},
  \bibinfo{journal}{Europhys. Lett.} \textbf{\bibinfo{volume}{81}},
  \bibinfo{pages}{40007} (\bibinfo{year}{2008}).

\bibitem[{\citenamefont{Jaatinen et~al.}(2009)\citenamefont{Jaatinen, Achim,
  Elder, and Ala-Nissila}}]{jaatinen-09}
\bibinfo{author}{\bibfnamefont{A.}~\bibnamefont{Jaatinen}},
  \bibinfo{author}{\bibfnamefont{C.}~\bibnamefont{Achim}},
  \bibinfo{author}{\bibfnamefont{K.~R.} \bibnamefont{Elder}}, \bibnamefont{and}
  \bibinfo{author}{\bibfnamefont{T.}~\bibnamefont{Ala-Nissila}},
  \bibinfo{journal}{Phys. Rev. E} \textbf{\bibinfo{volume}{80}},
  \bibinfo{pages}{031602} (\bibinfo{year}{2009}).

\bibitem[{\citenamefont{Jaatinen and Ala-Nissila}(2010)}]{jaatinen-10}
\bibinfo{author}{\bibfnamefont{A.}~\bibnamefont{Jaatinen}} \bibnamefont{and}
  \bibinfo{author}{\bibfnamefont{T.}~\bibnamefont{Ala-Nissila}},
  \bibinfo{journal}{J. Phys. Condens. Matter} \textbf{\bibinfo{volume}{22}},
  \bibinfo{pages}{205402} (\bibinfo{year}{2010}).

\bibitem[{\citenamefont{Greenwood et~al.}(2010)\citenamefont{Greenwood,
  Provatas, and Rottler}}]{greenwood-10}
\bibinfo{author}{\bibfnamefont{M.}~\bibnamefont{Greenwood}},
  \bibinfo{author}{\bibfnamefont{N.}~\bibnamefont{Provatas}}, \bibnamefont{and}
  \bibinfo{author}{\bibfnamefont{J.}~\bibnamefont{Rottler}},
  \bibinfo{journal}{Phys. Rev. Lett.} \textbf{\bibinfo{volume}{105}},
  \bibinfo{pages}{045702} (\bibinfo{year}{2010}).

\bibitem[{\citenamefont{Greenwood et~al.}(2011)\citenamefont{Greenwood,
  Rottler, and Provatas}}]{greenwood-11}
\bibinfo{author}{\bibfnamefont{M.}~\bibnamefont{Greenwood}},
  \bibinfo{author}{\bibfnamefont{J.}~\bibnamefont{Rottler}}, \bibnamefont{and}
  \bibinfo{author}{\bibfnamefont{N.}~\bibnamefont{Provatas}},
  \bibinfo{journal}{Phys. Rev. E} \textbf{\bibinfo{volume}{83}},
  \bibinfo{pages}{031601} (\bibinfo{year}{2011}).

\bibitem[{\citenamefont{Berry et~al.}(2012)\citenamefont{Berry, Provatas,
  Rottler, and Sinclair}}]{berry-12}
\bibinfo{author}{\bibfnamefont{J.}~\bibnamefont{Berry}},
  \bibinfo{author}{\bibfnamefont{N.}~\bibnamefont{Provatas}},
  \bibinfo{author}{\bibfnamefont{J.}~\bibnamefont{Rottler}}, \bibnamefont{and}
  \bibinfo{author}{\bibfnamefont{C.~W.} \bibnamefont{Sinclair}},
  \bibinfo{journal}{Phys. Rev. B} \textbf{\bibinfo{volume}{86}},
  \bibinfo{pages}{224112} (\bibinfo{year}{2012}).

\bibitem[{\citenamefont{Fallah et~al.}(2012)\citenamefont{Fallah, Stolle,
  Ofori-Opoku, Esmaeili, and Provatas}}]{fallah-12}
\bibinfo{author}{\bibfnamefont{V.}~\bibnamefont{Fallah}},
  \bibinfo{author}{\bibfnamefont{J.}~\bibnamefont{Stolle}},
  \bibinfo{author}{\bibfnamefont{N.}~\bibnamefont{Ofori-Opoku}},
  \bibinfo{author}{\bibfnamefont{S.}~\bibnamefont{Esmaeili}}, \bibnamefont{and}
  \bibinfo{author}{\bibfnamefont{N.}~\bibnamefont{Provatas}},
  \bibinfo{journal}{Phys. Rev. B} \textbf{\bibinfo{volume}{86}},
  \bibinfo{pages}{134112} (\bibinfo{year}{2012}).

\bibitem[{\citenamefont{Provatas and Elder}(2010)}]{nik-ken}
\bibinfo{author}{\bibfnamefont{N.}~\bibnamefont{Provatas}} \bibnamefont{and}
  \bibinfo{author}{\bibfnamefont{K.}~\bibnamefont{Elder}},
  \emph{\bibinfo{title}{Phase Field Methods in Material Science and
  Engineering}} (\bibinfo{publisher}{Wiley-VCH}, \bibinfo{year}{2010}).

\bibitem[{\citenamefont{Chaikin and Lubensky}(2000)}]{chaikin-lubensky}
\bibinfo{author}{\bibfnamefont{P.~M.} \bibnamefont{Chaikin}} \bibnamefont{and}
  \bibinfo{author}{\bibfnamefont{T.~C.} \bibnamefont{Lubensky}},
  \emph{\bibinfo{title}{Principles of Condensed Matter Physics}}
  (\bibinfo{publisher}{Cambridge University Press, Cambridge},
  \bibinfo{year}{2000}).

\bibitem[{\citenamefont{Herlach et~al.}(1998)\citenamefont{Herlach, Buhrer,
  Herlach, Maier, Notthoff, Platzek, and Reske}}]{herlach-98}
\bibinfo{author}{\bibfnamefont{D.}~\bibnamefont{Herlach}},
  \bibinfo{author}{\bibfnamefont{C.}~\bibnamefont{Buhrer}},
  \bibinfo{author}{\bibfnamefont{D.~M.} \bibnamefont{Herlach}},
  \bibinfo{author}{\bibfnamefont{K.}~\bibnamefont{Maier}},
  \bibinfo{author}{\bibfnamefont{C.}~\bibnamefont{Notthoff}},
  \bibinfo{author}{\bibfnamefont{D.}~\bibnamefont{Platzek}}, \bibnamefont{and}
  \bibinfo{author}{\bibfnamefont{J.}~\bibnamefont{Reske}},
  \bibinfo{journal}{Europhys. Lett.} \textbf{\bibinfo{volume}{44}},
  \bibinfo{pages}{98} (\bibinfo{year}{1998}).

\bibitem[{\citenamefont{Herlach}(2001)}]{herlach-01}
\bibinfo{author}{\bibfnamefont{D.~M.} \bibnamefont{Herlach}},
  \bibinfo{journal}{J. Phys.: Condens. Matter} \textbf{\bibinfo{volume}{13}},
  \bibinfo{pages}{7737} (\bibinfo{year}{2001}).

\bibitem[{\citenamefont{Holland-Moritz
  et~al.}(2003)\citenamefont{Holland-Moritz, Herlach, and
  Spaepen}}]{herlach-07}
\bibinfo{author}{\bibfnamefont{D.}~\bibnamefont{Holland-Moritz}},
  \bibinfo{author}{\bibfnamefont{D.~M.} \bibnamefont{Herlach}},
  \bibnamefont{and} \bibinfo{author}{\bibfnamefont{F.}~\bibnamefont{Spaepen}},
  \bibinfo{journal}{Superlattice Microst.} \textbf{\bibinfo{volume}{41}},
  \bibinfo{pages}{196} (\bibinfo{year}{2003}).

\bibitem[{\citenamefont{Brown}(1958)}]{magnetostrictive}
\bibinfo{author}{\bibfnamefont{W.~F.} \bibnamefont{Brown}}, in
  \emph{\bibinfo{booktitle}{Handbook of Physics}}, edited by
  \bibinfo{editor}{\bibfnamefont{E.~U.} \bibnamefont{Condon}} \bibnamefont{and}
  \bibinfo{editor}{\bibfnamefont{H.}~\bibnamefont{Odishaw}}
  (\bibinfo{publisher}{McGraw-Hill, New York}, \bibinfo{year}{1958}),
  chap.~\bibinfo{chapter}{8}.

\bibitem[{\citenamefont{Jackson}(1998)}]{jackson}
\bibinfo{author}{\bibfnamefont{J.~D.} \bibnamefont{Jackson}},
  \emph{\bibinfo{title}{Classical Electrodynamics}} (\bibinfo{publisher}{John
  Wiley and Sons Inc., New York}, \bibinfo{year}{1998}).

\bibitem[{\citenamefont{Goldenfeld et~al.}(2005)\citenamefont{Goldenfeld,
  Athreya, and Dantzig}}]{goldenfeld-05}
\bibinfo{author}{\bibfnamefont{N.}~\bibnamefont{Goldenfeld}},
  \bibinfo{author}{\bibfnamefont{B.~P.} \bibnamefont{Athreya}},
  \bibnamefont{and} \bibinfo{author}{\bibfnamefont{J.~A.}
  \bibnamefont{Dantzig}}, \bibinfo{journal}{Phys. Rev. E}
  \textbf{\bibinfo{volume}{72}}, \bibinfo{pages}{020601}
  (\bibinfo{year}{2005}).

\bibitem[{\citenamefont{Athreya et~al.}(2006)\citenamefont{Athreya, Goldenfeld,
  and Dantzig}}]{goldenfeld-06}
\bibinfo{author}{\bibfnamefont{B.~P.} \bibnamefont{Athreya}},
  \bibinfo{author}{\bibfnamefont{N.}~\bibnamefont{Goldenfeld}},
  \bibnamefont{and} \bibinfo{author}{\bibfnamefont{J.~A.}
  \bibnamefont{Dantzig}}, \bibinfo{journal}{Phys. Rev. E}
  \textbf{\bibinfo{volume}{74}}, \bibinfo{pages}{011601}
  (\bibinfo{year}{2006}).

\bibitem[{\citenamefont{Chan and Goldenfeld}(2009)}]{goldenfeld-09}
\bibinfo{author}{\bibfnamefont{P.~Y.} \bibnamefont{Chan}} \bibnamefont{and}
  \bibinfo{author}{\bibfnamefont{N.}~\bibnamefont{Goldenfeld}},
  \bibinfo{journal}{Phys. Rev. E} \textbf{\bibinfo{volume}{80}},
  \bibinfo{pages}{065105} (\bibinfo{year}{2009}).

\bibitem[{\citenamefont{Elder et~al.}(2012)\citenamefont{Elder, Huang, and
  Provatas}}]{elder-10}
\bibinfo{author}{\bibfnamefont{K.~R.} \bibnamefont{Elder}},
  \bibinfo{author}{\bibfnamefont{Z.-F.} \bibnamefont{Huang}}, \bibnamefont{and}
  \bibinfo{author}{\bibfnamefont{N.}~\bibnamefont{Provatas}},
  \bibinfo{journal}{Phys. Rev. E} \textbf{\bibinfo{volume}{81}},
  \bibinfo{pages}{011602} (\bibinfo{year}{2012}).

\bibitem[{\citenamefont{Landau et~al.}(2007)\citenamefont{Landau, Lifshitz, and
  Pitaevskii}}]{landau-lifshitz-electrodynamics}
\bibinfo{author}{\bibfnamefont{L.~D.} \bibnamefont{Landau}},
  \bibinfo{author}{\bibfnamefont{E.~M.} \bibnamefont{Lifshitz}},
  \bibnamefont{and} \bibinfo{author}{\bibfnamefont{L.~P.}
  \bibnamefont{Pitaevskii}}, \emph{\bibinfo{title}{Electrodynamics of
  Continuous Media}}, vol.~\bibinfo{volume}{8} (\bibinfo{publisher}{Elsevier
  Pte Ltd., Singapore}, \bibinfo{year}{2007}).

\bibitem[{\citenamefont{Atkin and Fox}(1980)}]{atkin-elesticity}
\bibinfo{author}{\bibfnamefont{R.}~\bibnamefont{Atkin}} \bibnamefont{and}
  \bibinfo{author}{\bibfnamefont{N.}~\bibnamefont{Fox}},
  \emph{\bibinfo{title}{An Introduction to the Theory of Elasticity}}
  (\bibinfo{publisher}{Longman, New York}, \bibinfo{year}{1980}).

\bibitem[{\citenamefont{Hohenberg and Halperin}(1977)}]{hohenberg-77}
\bibinfo{author}{\bibfnamefont{P.~C.} \bibnamefont{Hohenberg}}
  \bibnamefont{and} \bibinfo{author}{\bibfnamefont{B.~I.}
  \bibnamefont{Halperin}}, \bibinfo{journal}{Rev. Mod. Phys.}
  \textbf{\bibinfo{volume}{49}}, \bibinfo{pages}{435} (\bibinfo{year}{1977}).

\bibitem[{\citenamefont{Patra and Karttunen}(2005)}]{patra-05}
\bibinfo{author}{\bibfnamefont{M.}~\bibnamefont{Patra}} \bibnamefont{and}
  \bibinfo{author}{\bibfnamefont{M.}~\bibnamefont{Karttunen}},
  \bibinfo{journal}{Num. Meth. Part. Diff. Eqs} \textbf{\bibinfo{volume}{22}},
  \bibinfo{pages}{936} (\bibinfo{year}{2005}).

\bibitem[{\citenamefont{Read and Shockley}(1950)}]{read-shockley}
\bibinfo{author}{\bibfnamefont{W.~T.} \bibnamefont{Read}} \bibnamefont{and}
  \bibinfo{author}{\bibfnamefont{W.}~\bibnamefont{Shockley}},
  \bibinfo{journal}{Phys. Rev.} \textbf{\bibinfo{volume}{78}},
  \bibinfo{pages}{275} (\bibinfo{year}{1950}).

\bibitem[{\citenamefont{Herzer}(1990)}]{herzer-90}
\bibinfo{author}{\bibfnamefont{G.}~\bibnamefont{Herzer}},
  \bibinfo{journal}{IEEE Trans. Magn.} \textbf{\bibinfo{volume}{26}},
  \bibinfo{pages}{1397} (\bibinfo{year}{1990}).

\bibitem[{\citenamefont{Iwasaki et~al.}(2013)\citenamefont{Iwasaki,
  Masahito~Mochizuki, and Nagaosa}}]{iwasaki2013}
\bibinfo{author}{\bibfnamefont{J.}~\bibnamefont{Iwasaki}},
  \bibinfo{author}{\bibfnamefont{M.}~\bibnamefont{Masahito~Mochizuki}},
  \bibnamefont{and} \bibinfo{author}{\bibfnamefont{N.}~\bibnamefont{Nagaosa}},
  \bibinfo{journal}{Nature Comms.} \textbf{\bibinfo{volume}{4}},
  \bibinfo{pages}{1463} (\bibinfo{year}{2013}).

\bibitem[{\citenamefont{Chimata et~al.}(2012)\citenamefont{Chimata, Bergman,
  Bergqvist, Sanyal, and Eriksson}}]{chimata}
\bibinfo{author}{\bibfnamefont{R.}~\bibnamefont{Chimata}},
  \bibinfo{author}{\bibfnamefont{A.}~\bibnamefont{Bergman}},
  \bibinfo{author}{\bibfnamefont{L.}~\bibnamefont{Bergqvist}},
  \bibinfo{author}{\bibfnamefont{B.}~\bibnamefont{Sanyal}}, \bibnamefont{and}
  \bibinfo{author}{\bibfnamefont{O.}~\bibnamefont{Eriksson}},
  \bibinfo{journal}{Phys. Rev. Lett.} \textbf{\bibinfo{volume}{109}},
  \bibinfo{pages}{157201} (\bibinfo{year}{2012}).

\bibitem[{\citenamefont{Fal et~al.}(2013)\citenamefont{Fal, Plumer, Whithead,
  Mercer, {van Ek}, and Srinivasan}}]{fal2013}
\bibinfo{author}{\bibfnamefont{T.}~\bibnamefont{Fal}},
  \bibinfo{author}{\bibfnamefont{M.}~\bibnamefont{Plumer}},
  \bibinfo{author}{\bibfnamefont{J.}~\bibnamefont{Whithead}},
  \bibinfo{author}{\bibfnamefont{J.}~\bibnamefont{Mercer}},
  \bibinfo{author}{\bibfnamefont{J.}~\bibnamefont{{van Ek}}}, \bibnamefont{and}
  \bibinfo{author}{\bibfnamefont{K.}~\bibnamefont{Srinivasan}},
  \bibinfo{journal}{Appl. Phys. Lett.} \textbf{\bibinfo{volume}{102}},
  \bibinfo{pages}{202404} (\bibinfo{year}{2013}).

\bibitem[{\citenamefont{Kruzik and Prohl}(2006)}]{kruzik2006}
\bibinfo{author}{\bibfnamefont{M.}~\bibnamefont{Kruzik}} \bibnamefont{and}
  \bibinfo{author}{\bibfnamefont{A.}~\bibnamefont{Prohl}},
  \bibinfo{journal}{SIAM Rev.} \textbf{\bibinfo{volume}{48}},
  \bibinfo{pages}{439} (\bibinfo{year}{2006}).

\end{thebibliography}
\end{document}